\documentclass[11pt]{article}
 \usepackage{amsmath,amsfonts}
 \usepackage{graphicx}
 \usepackage{psfrag}
 \usepackage{flafter}
\def \beq  {\begin{equation}}
\def \eeq  {\end{equation}}
\def \beqar {\begin{eqnarray}}
\def \eeqar {\end{eqnarray}}
\def\sqr#1#2{{\vcenter{\vbox{\hrule height.#2pt
\hbox{\vrule width.#2pt height#1pt \kern#1pt
\vrule width.#2pt}\hrule height.#2pt}}}}

\def\L {{\cal L}}
\def\N {{\mathcal N}}

\def\vf {{\varphi}}

\def\dag {{\dagger}}

\def\hs {\hat{s}}
\def\hx {\hat{x}}

\def\dag {\dagger}

\def\del {\partial}

\def\a {\alpha}
\def\b {\beta}

\def\e {\epsilon}
\def\ve {\varepsilon}
\def\d {\delta}

\def\s {\sigma}
\def\l {\lambda}
\def\o {\omega}

\def\vf {{\varphi}}

\def \L {{\cal L}}
\def \M {{\cal M}}

\begin{document}
\def \CMP {{ Commun. Math. Phys.}}
\def \PRL {{ Phys. Rev. Lett.}}
\def \PL {{Phys. Lett.}}
\def \NPBProc {{ Nucl. Phys. B (Proc. Suppl.)}}
\def \NP {{ Nucl. Phys.}}
\def \RMP {{ Rev. Mod. Phys.}}
\def \JGP {{ J. Geom. Phys.}}
\def \CQG {{ Class. Quant. Grav.}}
\def \MPL {{Mod. Phys. Lett.}}
\def \IJMP {{ Int. J. Mod. Phys.}}
\def \JHEP {{ JHEP}}
\def \PR {{Phys. Rev.}}
\def \JMP {{J. Math. Phys.}}
\def\JoP {{J. Phys.}}
\begin{titlepage}
\null\vspace{-62pt} \pagestyle{empty}
\begin{center}
\rightline{CCNY-HEP-07/x} \rightline{March 2007} \vspace{1truein}
{\Large\bfseries

Gluon Radiation of an Expanding Color Skyrmion in the Quark-Gluon
Plasma
}\\
\vskip .2in\noindent

\vspace{.5in} {\bf\large Jian Dai}\footnote{E-mail:
\fontfamily{cmtt}\fontsize{11pt}{15pt}\selectfont
jdai@sci.ccny.cuny.edu}
\\
\vspace{.15in}{\itshape Physics Department\\
City College of the CUNY\\
New York, NY 10031}\\

\fontfamily{cmr}\fontsize{11pt}{15pt}\selectfont
 \vskip 1in
\centerline{\large\bf Abstract}
\end{center}

 The density of states and energy spectrum of the gluon radiation are
 calculated for the color current of an expanding hydrodynamic skyrmion
 in the quark gluon plasma with a semiclassical method. Results
 are compared with those in literatures.

\end{titlepage}
\pagestyle{plain} \setcounter{page}{1} \setcounter{footnote}{0}

\section{Introduction}

 In this letter, we address the issue of gluon radiation during the
 hydrodynamic stage in the evolution of the deconfined hot QCD
 matter or quark gluon plasma (QGP) \cite{expt1} (for review see for example \cite{expt2}).

 The medium induced gluon radiation has been thoroughly explored in
 the context of final state partonic energy loss or ``jet quenching'' \cite{KW}.
 The spatially extended nuclear matter affects the processes of fragmentation and hadronization
 of the hard partons produced in the relativistic heavy ion collisions.
 Essentially all high $p_\bot$ hadronic observables are affected
 at collider energies and the degree of the medium modification
 can give a characterization of the hot QCD matter in the deconfined phase.
 In principle, the medium induced radiation effect emerges from thermal QCD {\it per se}.
 However, in practice, different approximation schemes are applied giving consistent results \cite{SW,W}.
 On the other hand, gluon radiation has also been considered in the
 context of gluon density saturation in the initial stage, where a
 strongly interacting gluonic atmosphere is crucial for the
 rapid local thermalization for the deconfined QCD matter \cite{KR}.

 The time evolution of the RHIC ``fireball'' can influence the observable particle production spectra.
 Given a strong initial interaction, the resulting state of matter is usually
 modeled as a relativistic fluid undergoing a hydrodynamic flow.
 Generalized fluid mechanics that characterizes the long-distance physics of the transport of
 color charges has been developed for this purpose \cite{NFM} (for review see \cite{NFMR}).
 Recently, we discovered a type of single skyrmion solutions in color fluid \cite{DN}.
 Moreover, we found an interesting case in which the time-dependent skyrmion expands in
 time, which is in accordance with the expanding nature of the fireball generated
 in RHIC experiments \cite{D}.
 The pattern of gluon radiation pertaining to the color current of these non-static
 configurations is an important character of this color skyrmion.
 So in this letter we calculate this radiation spectrum in a semiclassical approach.
 The main results from our calculation are the following.
 There is a fast fall-off in the UV side of the spectrum
 but a smooth peak dominates the intermediate energy.
 And in IR, a long tail is the characteristic feature.

 The organization of this paper is the following.
 In Sect.~\ref{Current}, after a brief review of the nonabelian fluid mechanics,
 we calculate the nonabelian current corresponding to the soliton solution.
 In Sect.~\ref{Gluon}, semiclassical gluon radiation is calculated.
 In Sect.~\ref{COM}, comparison of the radiation spectrum in our hydrodynamic approach and in other approaches
 is carried out.

\section{Color current of an expanding soliton}\label{Current}

 Given the thermalization of hot QCD matter above the deconfinement transition temperature,
 the transport of the color charges in the volume of the nuclear
 size can be modeled by a nonlinear sigma model in a first-order formalism
 \beq\label{L}
  \L=j^\mu \o_\mu - F(n) - g_{eff}J^{a\mu}A^a_\mu.
 \eeq
 This nonlinear sigma model describes an ideal fluid system.
 The configuration of this fluid is described by a
 group element field $U$, which shows up in the velocity field $\o_\mu$
 \beq
  \o_\mu=-{i\over 2}Tr(\s_3 U^\dag \del_\mu U).
 \eeq
 Conjugate to the velocity is the abelian charge current $j^\mu$.
 It is easy to see that the first term in the lagrangian density (\ref{L}) gives rise to
 the canonical structure of the fluid system.
 The fact that we will consider only one abelian charge current
 means that $U$ takes value in an $SU(2)$ group.
 The information about the equation of state (EOS) of the fluid is contained in the second term, which
 is essentially the free energy density of the fluid.
 In fact, energy and pressure densities are given by the ideal fluid formula
 \beq\label{EOS0}
  \e=F,~~p=nF^\prime-F.
 \eeq
 Here $n$ is the invariant length of $j^\mu$, $n^2=j^\mu j_\mu$.
 The third term is the gauge coupling of the fluid with an external gluon field $A^a_\mu$ with
 an effective coupling $g_{eff}$.
 $J^{a\mu}$ is the nonabelian charge current
 which is related to the abelian current by the Eckart factorization $J^{a\mu}=Q^aj^\mu$
 where $Q^a$ is
 the nonabelian charge density of the fluid configuration
 \beq
  Q^a={1\over 2}Tr(\s_3U^\dag \s^aU).
 \eeq
 For $SU(2)$ group, $a=1,2,3$.

 When the temperature is relatively high, we approximate the EOS by
 \beq\label{EOS}
  \e = 3p
 \eeq
 which is known in relativistic fluid mechanics to describe radiation.
 As a result, the free energy
 density can be obtained by integrating Eq.~(\ref{EOS0}),
 \beq
  F={\b\over 4/3} n^{4/3}
 \eeq
 where $\b$ is a dimensionless constant of integration.
 In this case, and without an external gluon field, the fluid system
 in (\ref{L}) possesses a class of expanding soliton solutions which can be studied via variational and
 collective coordinate methods \cite{D}.
 \beq\label{S}
  U=U\Bigl({{\mathbf x}\over R(t)}\Bigr),~~R(t)\approx R_0({t\over
  \tau}+1)^{4/3}\theta(t)
 \eeq
 where
 $R_0$ and $\tau$ are the spacial and temporal characterizations of
 the variational soliton and $\theta(t)$ the usual step function in
 time direction. Physically, it is certainly very interesting to
 understand the origin of these two scales from a fundamental level.
 The approximation in (\ref{S}) is valid provided $\tau \ll R_0$.
 This condition enables us to define a small parameter
 \beq
  \l={\tau\over R_0}.
 \eeq
 For our purpose, we calculate the nonabelian current in (\ref{L})
 corresponding to the soliton solution in Eq.~(\ref{S}).
 To do so, the {\em hedgehog ansatz} is specified for the solution (\ref{S})
 \beq
  U=\cos\phi+i\s\cdot\hx \sin\phi
 \eeq
 where $\hx$ is the unit vector and $\phi$ is given by the stereographic map
 \beq
  \sin\phi={2s\over 1+s^2},~~
  \cos\phi=\pm {1-s^2\over 1+s^2}.
 \eeq
 We write $s$ as the dimensionless coordinate $x/R(t)$. The
 sign in the expression of $\cos\phi$ signifies a topological
 charge which is the {\em skyrmion number}. The negative sign gives the skyrmion
 number $+1$ or a skyrmion and the positive sign the skyrmion number is $-1$
 or an anti-skyrmion. We will take the positive sign in the following.
 By expressing the abelian current $j^\mu$ in terms of the velocity $\o_\mu$
 through the equation of motion, we derive the following expression for
 the nonabelian current
 \beqar\nonumber
  d^3xJ^{a\mu}&=&\Bigl({2\over\b}\Bigr)^3\cdot {d^3s\over (1+s^2)^6}\cdot (\hs_3^2s^2\dot{R}^2-1)\cdot\\
  &&\Bigl(
  \d^a_3(1-6s^2+s^4)+4\e^{a3b}\hs_bs(1-s^2)+8\hs_3\hs_as^2
  \Bigr)\cdot \left(\begin{array}{c}-\hs_3s(1+s^2)\dot{R}\\
  2\hs_1\hs_3s^2-2\hs_2s\\2\hs_2\hs_3s^2+2\hs_1s\\
  2\hs_3^2s^2-s^2+1
  \end{array}\right).\label{J}
 \eeqar
 The current in (\ref{J}) has a natural form of a multipole
 expansion due to the skyrmion orientation in the color space.
 In this letter we only consider the effect of the lowest mode and the effects of higher
 polarization will be considered elsewhere.
 The spherically
 symmetric part in the current is contained only in the third component
 \beq
  \Bigl(d^3xJ^{a3}\Bigr)_0=-\d^a_3\Bigl({2\over\b}\Bigr)^3{d^3s\over
  (1+s^2)^6}P_6(s)
 \eeq
 where $P_6(s)=1-7s^2+7s^4-s^6$.

\section{Semiclassical gluon radiation}\label{Gluon}

 Now we consider the interaction between the expanding color skyrmion and the
 hard partons. Since the transfer momentum between hard partons is in
 high order to that between hard parton and soliton,
 we expect a hierarchy between the
 partonic coupling $g_{YM}$ and the effective coupling $g_{eff}$.
 Accordingly, gluon self-interaction in terms like
 $F^a_{\mu\nu}F^{a\mu\nu}$ can be omitted so we can work with a free parton picture.
 Then the gauge coupling in (\ref{L}) becomes the coupling between a classical
 current and a free quantum field for gluon.
 In this approximation, the lowest order semiclassical amplitude is
 given by
 \beq\label{A}
  i\M=g_{eff}\langle 1|\int d^4x J^{a\mu}\hat{A}^a_\mu |0\rangle.
 \eeq
 $|0\rangle$ and $|1\rangle$ are gluonic Fock vacuum and
 one-gluon state. The gluon factor in (\ref{A}) is given by the wave function
 \beq
  \langle 1|\hat{A}^a_\mu(x) |0\rangle=\vf^a\ve_\mu {e^{ik\cdot
  x}\over \sqrt{2\o}}
 \eeq
 where the color and helicity parts $\vf$, $\ve$ will be summed over eventually.
 Putting the current in, we have
 \beq
  i\M=A(k)\int dt e^{i\o t}
  \int{d^3s\over (1+s^2)^6}e^{-iR(t){\mathbf k}\cdot{\mathbf
  s}}P_6(s)
 \eeq
 where $A(k)=-(2/\b)^3g_{eff}\vf^3\ve_3/\sqrt{2\o}$.
 The spatial Fourier transformation can be completed analytically
 \beq\label{F0}
  i\M=B(k)\int dt e^{i\o t-R(t)k}Q_4(R(t)k)
 \eeq
 where $B(k)=\pi^2A(k)/120$ and $Q_4(x)=5x^2-5x^3+x^4$.
 To go further, we need to specify $R(t)$ in this equation to the form given in (\ref{S}).
 This gives
 \beq
  i\M=B(k)e^{-i\o\tau}{\eta \over\o }\int\limits_{{\o\tau\over\eta}}^\infty
  dt e^{i\eta t-t^{4/3}}Q_4(t^{4/3})
 \eeq
 where $\eta=\o\tau/(kR_0)^{3/4}$. With onshell condition $\o=k$,
 $\eta=\l\kappa^{1/4}$ where $\kappa$ is defined to be $R_0k$.
 Accordingly,
 \beq
  i\M=\Bigl(-{\pi^2\over 15\sqrt{2}}\Bigr)\Bigl({g_{eff}\over\b^3}\l R_0^{3/2}\Bigr)\Bigl(\vf^3\ve_3 e^{-i\o\tau}\Bigr)
  \Bigl({i\widetilde{\M}_\l(\kappa)\over\kappa^{5/4}}\Bigr)
 \eeq
 where
 \beq
  i\widetilde{\M}_\l(\kappa)=\int\limits_{\kappa^{3/4}}^\infty
  dt e^{i\l\kappa^{1/4} t-t^{4/3}}Q_4(t^{4/3})
 \eeq

\begin{figure}[hbtp]
  \centering
  \includegraphics[width=.48\textwidth]{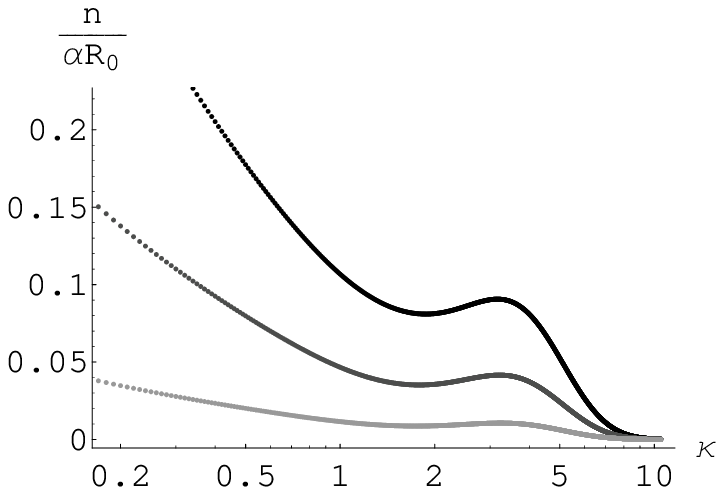}
  \hspace{.02\textwidth}
  \includegraphics[width=.48\textwidth]{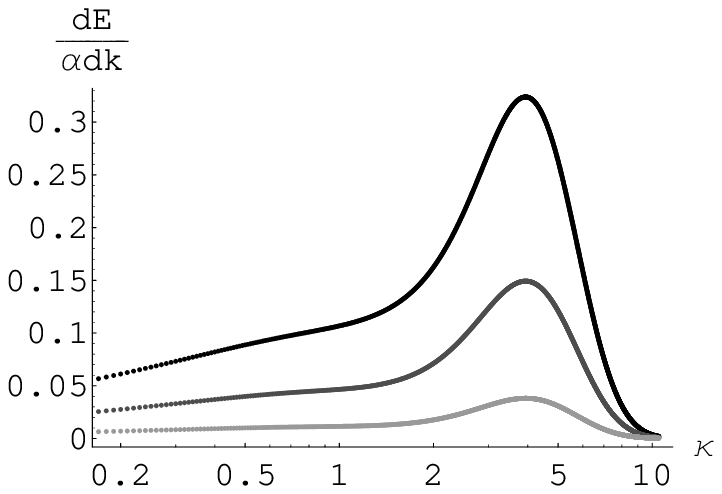}
  \caption{Density of states and energy spectrum for
  $\lambda=1/5$ (Black), $2/15$ (Deep Gray) and $1/15$ (Light Gray).}\label{DandS}
 \end{figure}

 The radiation spectrum is given by $dE=kd\N$. $E(k)$ is the total energy radiated over the entire time of
 expansion as a function of $k$.
 The number distribution is
 \beq
  d\N = \sum\limits_{c,h} |\M|^2d^3k
 \eeq
 where the summation is over colors and helicities of the
 gluon. In a spherically symmetric setting, $d\N=n dk$ where $n$ is the density of
 states
 \beq
  n=4\pi k^2\sum\limits_{c,h}|\M|^2.
 \eeq
 By straightforward calculation,
 \beqar
  n &=& \a R_0
  \lambda^2\kappa^{-1/2}|\widetilde{\M}_\l(\kappa)|^2,\\
  {dE\over dk}&=& \a \lambda^2
  \kappa^{1/2}|\widetilde{\M}_\l(\kappa)|^2.
 \eeqar
 where $\a\equiv (2\pi^5/225)(g_{eff}^2/\b^6)$.
 The numerical results for $\lambda=1/15,2/15,1/5$ are given in
 Fig. ~\ref{DandS}.

\section{Comparison and discussion}\label{COM}

 Understanding the pattern of gluon radiation in relativistic heavy
 ion collision processes is important for making an accurate determination of
 the physical mechanisms from the measurement of its decay products.

\begin{figure}[h]
  \centering
  \includegraphics[width=.80\textwidth]{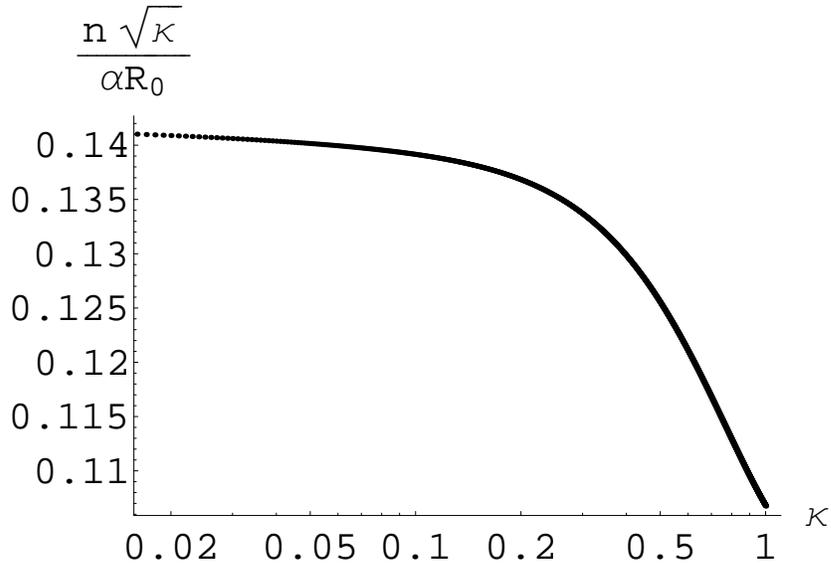}
  \caption{$n/(1/\sqrt{\kappa})$ in small $k$ for $\lambda=.2$}\label{Fig2}
 \end{figure}

 In \cite{KR}, the authors extracted the asymptotic behavior of the
 number density in small $k$ is of the $1/k$ form. In our case, the
 asymptotic of the number density in small $k$ is $\sim 1/\sqrt{k}$.
 (See Fig. ~\ref{Fig2}.)
 The difference comes from the fact that the medium size is taken to be
 infinitely large in \cite{KR} while in our case the medium size is
 characterized by the soliton size $R_0$. So the IR behavior in our
 case is softer.

 For the case of jet quenching, the radiation energy lost is due to scattering off the hard quarks.
 A popular approach is to model the medium as a collection of colored
 static scattering centers \cite{GW}. This approach can be
 extended to the expanding medium \cite{SW} though the gluon radiation by the expanding medium itself
 is not included. In fact, the medium induced gluon radiation is
 characterized by the frequency
 \beq
  \o_C={1\over 2}\hat{q}L^2
 \eeq
 where $\hat{q}$ is the quenching parameter, estimated to be $.04 \sim .16 GeV^2/fm$, and $L$ is the in-medium
 path length of a hard parton \cite{GVWZ}. In general $\o_C$ is significantly
 larger than the characteristic momentum in our case $1/R_0$. So
 there is a hierarchy between the medium induced gluon radiation
 spectrum and the gluon radiation spectrum by the medium.

 Our hydrodynamical approach opens up another interesting
 possibility to address the eccentricity of the elliptic flow either
 intrinsically by considering the nonabelian color current or
 exogenously by considering the gluon radiation patterns. This will
 be the topic of the follow-up to this work.

\vskip .2in\noindent

{\bf Acknowledgment}. This work was supported by a CUNY
Collaborative Research Incentive grant. The author has greatly
benefited from the mentoring by V. P. Nair.


\begin{thebibliography}{99}
 \bibitem{expt1} PHENIX Collaboration, K. Adcox, {\it et al}, Nucl. Phys. {\bf A757} (2005) 184-283, nucl-ex/0410003;
  I. Arsene {\it et al}. BRAHMS collaboration, Nucl. Phys. {\bf A757} (2005) 1-27, nucl-ex/0410020;
  B. B. Back {\it et al} (PHOBOS), Nucl. Phys. {\bf A757} (2005) 28-101, nucl-ex/0410022;
  STAR Collaboration: J. Adams, {\it et al}, Nucl. Phys. {\bf A757} (2005) 102-183, nucl-ex/0501009.
 \bibitem{expt2} Berndt Muller, James L. Nagle, nucl-th/0602029.
 \bibitem{KW} Alexander Kovner, Urs A. Wiedemann, ``Gluon Radiation and Parton Energy Loss'',
  in {\em Quark Gluon Plasma 3} Editors: R. C. Hwa and X. Wang World
  Scientific Singapore, hep-ph/0304151.
 \bibitem{SW} Carlos A. Salgado, Urs Achim Wiedemann, Phys. Rev. {\bf D68} (2003) 014008, hep-ph/0302184.
 \bibitem{W} Urs A. Wiedemann, Nucl. Phys. {\bf B588} (2000) 303,
 hep-ph/0005129.
 \bibitem{KR} Yuri V. Kovchegov, Dirk H. Rischke, Phys. Rev. {\bf C56} (1997) 1084, hep-ph/9704201.
 \bibitem{NFM} R. Jackiw, V.P. Nair, So-Young Pi,
 Phys. Rev. {\bf D62} (2000) 085018, hep-th/0004084;
 B. Bistrovic, R. Jackiw, H. Li, V.P. Nair, S.-Y. Pi, Phys. Rev. {\bf D67} (2003)
 025013, hep-th/0210143.
 \bibitem{NFMR} R. Jackiw, V.P. Nair, S.-Y. Pi, A.P. Polychronakos,
 \JoP {\it A. Math. Gen.} {\bf 37} (2004) R327.
 \bibitem{DN} Jian Dai, V.P. Nair, Phys. Rev. {\bf D74} (2006) 085014, hep-ph/0605090.
 \bibitem{D} Jian Dai, ``Stability and Evolution of Color Skyrmions in the Quark-Gluon Plasma'',
 hep-ph/0612260.
 \bibitem{GW} M. Gyulassy, X. Wang, Nucl. Phys. {\bf B420} (1994) 583.
 \bibitem{GVWZ} Miklos Gyulassy, Ivan Vitev, Xin-Nian Wang, Ben-Wei
 Zhang, ``Jet Quenching and Radiative Energy Loss in Dense Nuclear
 Matter'', in {\em Quark Gluon Plasma 3} Editors: R. C. Hwa and X. Wang World Scientific Singapore,
 nucl-th/0302077.
 \end{thebibliography}
\end{document}